\documentclass[twocolumn,showpacs,preprintnumbers,amsmath,amssymb]{revtex4-1}

\usepackage{graphicx}
\usepackage{dcolumn}
\usepackage{bm}
\usepackage{color}
\newcommand{\sqrtsnn}{\sqrt{s_{\mbox{\tiny{\it{NN}}}}}}

\newcommand{\hijing}{{\sc hijing} }
\newcommand{\hydjet}{{\sc hydjet}}
\newcommand{\pyquen}{{\sc pyquen} }
\newcommand{\fastmc}{{\sc fastmc} }
\newcommand{\ampt}{{\sc ampt} }
\newcommand{\epos}{{\sc epos} }
\newcommand{\qgsjet}{{\sc qgsjet} }
\newcommand{\paciae}{{\sc paciae} }
\newcommand{\angantyr}{{\sc angantyr} }

\newcommand{\GeVc}{${\rm GeV}/c$ }

\begin{document}


\title{Toward a description of the centrality dependence of the charge 
balance function in the HYDJET++ model}

\author{ A.S.~Chernyshov$^1$, G.Kh.~Eyyubova$^1$, V.L.~Korotkikh$^{1}$, 
I.P.~Lokhtin$^1$, L.V.~Malinina$^1$, S.V.~Petrushanko$^1$, 
A.M.~Snigirev$^{1,2}$, and E.E.~Zabrodin$^{1,3}$}

\affiliation{$^{1}$ Skobeltsyn Institute of Nuclear Physics, Lomonosov 
Moscow State University, RU-119991 Moscow, Russia }

\affiliation{$^{2}$ Bogoliubov Laboratory of Theoretical Physics, JINR, 
RU-141980 Dubna, Russia }

\affiliation{$^{3}$ Department of Physics, University of Oslo, 
N-0316 Oslo, Norway}

\begin{abstract}
Data from the Large Hadron Collider on the charge balance function in 
Pb+Pb collisions at center-of-mass energy 2.76~TeV per nucleon pair are 
analyzed and interpreted within the framework of the \hydjet++ model. 
This model allows us to qualitatively reproduce the experimentally 
observed centrality dependence of the balance function widths at 
relatively low transverse momentum intervals due to the different 
charge creation mechanisms in soft and hard processes. However, a fully 
adequate description of the balance function in these intervals implies 
an essential modification of the model by including exact charge 
conservation via the canonical rather than the grand canonical ensemble. 
A procedure is proposed for introducing charge correlations into the 
thermal model without changing other model parameters. With increasing 
transverse momenta, the default model results describe the experimental 
data much better because the contribution of the soft component of the 
model is significantly reduced in these transverse momentum intervals. 
In practical terms, there is a transition to a single source of charge 
correlations, namely, charge correlations in jets in which exact charge 
conservation holds at each stage.

{\it Keywords:} relativistic heavy-ion collisions, charge balance 
function, soft and hard processes, canonical charge conservation
\end{abstract}
\pacs{25.75.-q, 24.10.Nz, 24.10.Pa}

\maketitle


\section{Introduction}
\label{intro}

With the Relativistic Heavy Ion Collider (RHIC) and the Large Hadron 
Collider (LHC) in operation, a number of exquisite and intriguing 
phenomena have been revealed that could have never been systematically 
studied ith the previous generation of accelerators. Among these, 
anisotropic flow, the energy loss of high transverse momentum 
particles, and the charge balance function may be used as sensitive 
probes of the collective properties of a new state of matter, 
quark-gluon plasma (QGP); see, e.g., \cite{qm19,Proceedings:2019drx}. 
The large number of these physical observables measured in heavy-ion 
collisions during RHIC and LHC operation can be successfully described 
within the framework of the popular \hydjet++ model 
\cite{HYDJET_manual}. Calculations applying this model, such as the 
transverse momentum spectra, pseudorapidity and centrality dependence 
of inclusive charged particle multiplicity, and $\pi^\pm \pi^\pm$ 
correlation radii in central Pb+Pb collisions \cite{Lokhtin:2012re}, 
the centrality and momentum  dependence of second and higher-order 
harmonic coefficients \cite{Bravina:2013xla}, flow fluctuations 
\cite{Bravina:2015sda}, jet quenching effects \cite{jet1,jet2}, and 
angular dihadron correlations \cite{Eyyubova:2014dha} are in fair 
agreement with the experimental data. 

Some of the experimental measurements proposed as indicators of 
QGP creation involve an implicit understanding of quark production 
dependence on time. While, for example, the fact of strangeness 
enhancement has been well established indeed, it remains unclear 
whether the arbitrary mechanism is dominant during early ($\tau < 
1$~fm/$c$) or late stages of the QGP fireball expansion. The 
charge balance function (CBF) was proposed in \cite{Bass} as a means 
for pinpointing the time of quark production by quantifying the
separation of balancing charges. Connections between the CBFs and 
charge fluctuations and correlations were discussed in \cite{JP_02}, 
whereas in \cite{pratt2} the possibility was demonstrating of 
applying species-binned CBFs as constraints on the diffusivity. 
Emergence of the charge balance functions has been studied, e.g., 
in hydrodynamics \cite{LSS_13}, in a thermal model with resonances
\cite{FBB_04}, and in a coalescence model \cite{Bialas_04}, as well 
as in other dynamic and statistical models \cite{Ch_04}.
Experimentally, the charge balance function has been measured at 
RHIC~\cite{star_03,star_10,star_16} in $pp$, $d+$Au, and Au+Au 
collisions and at LHC \cite{alice_epjc,alice_2013,alice_jpan} in 
$pp$, $p+$Pb, and Pb+Pb collisions. The charge balance function 
dependence on the collision centrality and beam energy are 
presented in these studies alongside some experimental techniques. 
An extensive overview of the CBF from both theoretical and 
experimental perspectives can be found in \cite{tawfik}. In the 
present paper we focus on the charge balance function as a source 
of valuable insights into the charge creation mechanism as well.

The width of the balance function reported by the STAR Collaboration 
in \cite{star_03} and by the ALICE Collaboration 
in \cite{alice_2013} decreases with increasing centrality of the 
collisions. This centrality dependence is not reproduced by many 
event generators, e.g., \hijing~\cite{hijing1,hijing2} and 
\ampt~\cite{ampt1,ampt2,ampt3}, and poses a challenge for \hydjet++ 
as well since a majority of soft particles is produced in the grand 
canonical ensemble approach, where charge conservation holds in the
mean only. Nevertheless, a reasonable simple modification of the 
current version of the \hydjet++ model allows us to reproduce the 
experimentally observed nontrivial centrality dependence of the 
balance function widths qualitatively due to the different charge 
creation mechanisms in soft and hard processes. Further essential 
modernizations have also been suggested within the canonical ensemble 
approach to describe the charge balance function quantitatively.

The paper is organized as follows. Basic characteristics of the model 
are sketched in Sec.~\ref{sec_2}. Section~\ref{sec_3} presents a
comparison of model results calculated within the default version of
\hydjet++ with the experimental data. The widths of the CBFs in the 
model are broader than the experimental ones. Possible variations in 
the partial contributions of soft and hard processes to the CBFs are 
studied. A modification procedure that takes into account exact charge 
conservation in a single event without changing its bulk 
characteristics is introduced in Sec.~\ref{sec_4}. Implementation of 
this procedure in \hydjet++ allows us to reproduce both the widths and 
the centrality dependence of the charge balance functions fairly well 
compared to the data. Conclusions are drawn in Sec.~\ref{sec_5}.

\section{The HYDJET++ model}
\label{sec_2}

It should be noted that currently there are many competing Monte Carlo 
event generators successfully describing the soft and hard momentum 
components of particle production in ultrarelativistic nuclear 
collisions separately. The \hydjet++ model is one of the few, such as
\epos \cite{epos}, \qgsjet \cite{qgsjet}, \paciae 
\cite{paciae_1,paciae_2}, \angantyr \cite{angantyr}, that aim to treat 
the soft and hard physics of the collisions simultaneously. For 
instance, the presence of both soft and hard processes ``in one package'' 
has recently allowed the model to reproduce \cite{PhysRevC103} the 
experimentally observed \cite{cms2018} nontrivial centrality dependence 
of elliptic flow correlations at low and high transverse momenta in 
Pb+Pb collisions at LHC energies. The origin of the correlations 
between the low and high-$p_T$ flow components in (semi)peripheral 
Pb+Pb collisions was traced to the correlations of particles in jets. 
As we will see below, these correlations are also important for the 
description of the charge balance function. Details of the model can be 
found in the \hydjet++ manual \cite{HYDJET_manual}. Here we stress and 
recall briefly only the main features crucial for the present study. 

The \hydjet++ is a Monte Carlo event generator for the simulation of 
relativistic heavy-ion collisions considered as a superposition of two 
independent components, namely, the soft hydro-type state and the hard 
state arising from in-medium multiparton fragmentation. It is based 
on the adapted version of the event generator 
\fastmc~\cite{Amelin:2006qe,Amelin:2007ic} and the \pyquen partonic 
energy loss model~\cite{Lokhtin:2005px}.

In the hard part the partons propagate through the expanding 
quark-gluon plasma and lose energy due to parton rescattering and 
gluon radiation. A number of jets is generated in accordance with a 
binomial distribution. The mean number of jets produced in an $A+A$ 
event is a product of the number of nucleon-nucleon ($NN$) binary 
subcollisions at a given impact parameter $b$ and the integral cross 
section of the hard subprocess in these subcollisions with the minimum 
transverse momentum transfer $p_T^{\rm min}$. Its value is one of the 
input key parameters of the model, because partons produced in 
(semi)hard processes with the momentum transfer lower than 
$p_T^{\rm min}$ are excluded from the hard process treatment. Their 
hadronization products are automatically added to the soft component 
of the particle spectrum.

The soft part of the model is represented by the thermalized hadronic 
state generated on the chemical and thermal freeze-out hypersurfaces 
prescribed by the parametrization of relativistic hydrodynamics with 
preset freeze-out conditions. Particle multiplicities are calculated 
within the effective thermal volume using a statistical model approach. 
The effective volume absorbs the collective velocity profile and the 
hypersurface shape and cancels out in all particle number ratios. 
Therefore, the latter do not depend on the freeze-out details so long 
as the local thermodynamic parameters are independent of spatial 
coordinates \cite{Amelin:2006qe,Amelin:2007ic}. The concept of the 
effective volume is applied to calculate the hadronic composition at 
both chemical and thermal freeze-outs. The number of particles in an 
event is calculated according to a Poisson distribution around its 
mean value, which is supposed to be proportional to the number of 
participating nucleons for a given impact parameter of A+A collision. 
To simulate the elliptic and triangular flow effects, a hydro-inspired 
parametrization \cite{Wiedemann:1997cr} for the momentum 
and spatial anisotropy of soft hadron emission source is implemented; 
see \cite{HYDJET_manual,Crkovska:2016flo,Bravina:2017rkp} for details. 
In the following we refer to \hydjet++ version 2.4, freely available 
for download at \cite{version_2.4}, as the default version.

\section{CBF and contributions from soft and hard processes}
\label{sec_3}

The charge balance function has been proposed as a convenient measure 
of the correlation between oppositely charged particles \cite{Bass}. 
It provides valuable insight into the charge creation mechanism and 
can address the fundamental question concerning the hadronization 
process in nuclear collisions at relativistic energies. The final 
degree of correlations is reflected in the balance function and 
consequently in its width. It is defined as 
\begin{eqnarray}
\displaystyle
\label{bf}
B(\Delta \eta) &=& \frac{1}{2}  \Big[\frac{\langle N_{+-}(\Delta \eta)
   \rangle - \langle N_{++}(\Delta \eta)\rangle }{\langle N_{+}
   \rangle } \nonumber \\
&+&\frac{\langle N_{-+}(\Delta \eta)\rangle - \langle N_{--}(\Delta \eta)
   \rangle }{\langle N_{-} \rangle }\Big]\ ,
\end{eqnarray}
where $\langle N_{+-}(\Delta \eta) \rangle$ is the average number of 
opposite-charge pairs with particles separated by the relative 
pseudorapidity $\Delta \eta =\eta_{+}-\eta_{-}$, and similarly for 
$\langle N_{-+}(\Delta \eta) \rangle$, $\langle N_{++}(\Delta \eta)
\rangle$ and $\langle N_{--}(\Delta \eta) \rangle $. Both particles of 
the pair have to fall within a certain pseudorapidity interval, for 
instance $|\eta|< 0.8$, in accordance to the ALICE analysis conditions 
\cite{alice_2013}. $\langle N_{+} \rangle$ (and $\langle N_{-} 
\rangle $) is the number of positive (negative) charge particles in 
the pseudorapidity interval $|\eta|< 0.8$ averaged over all events. 
The charge balance function $B(\Delta\varphi)$ as a function of the 
relative azimuthal angle $\Delta\varphi$ is defined similarly. Each 
term $N(\Delta\eta)$ is corrected for the acceptance limitation, 
reflecting the fact that the number of pairs in the limited acceptance 
has a maximum at $\Delta\eta = 0$.

\begin{figure}[htpb]
\includegraphics[width=0.5\textwidth]{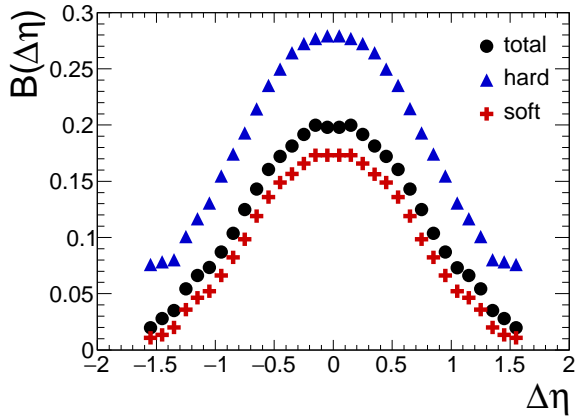}
	\caption{(Color online)
Balance function as a function of $\Delta\eta$ in Pb+Pb 
collisions at $\sqrtsnn = 2.76$~TeV with centrality 0-5\% calculated 
in \hydjet++ for each component, soft (crosses) and hard (triangles), 
together with the resulting total values (circles).}
\label{fig_alice1}
\end{figure}

The width of the balance function distribution is defined as
\begin{eqnarray}
\displaystyle
\label{bf-w}
\langle \Delta \eta \rangle =\sum\limits_{i = 1}^k [B(\Delta \eta_{i}) 
        \cdot \Delta\eta_{i}]/\sum\limits_{i = 1}^k B(\Delta \eta_{i})\ ,
\end{eqnarray}
where $B(\Delta \eta_{i})$ is the balance function value for each 
bin $\Delta\eta_{i}$, with the sum running over all $k$ bins. 

Figure~\ref{fig_alice1} shows the charge balance function in terms 
of the pseudorapidity $\Delta\eta$ calculated in the \hydjet++ model 
separately for each component, soft and hard, together with the 
resulting total value. Calculations are done for Pb+Pb collisions at
center-of-mass energy per nucleon pair $\sqrtsnn = 2.76$~TeV with 
centrality 0--5\%. One can see that the resulting total balance 
function is mainly dominated by the balance function of the soft 
component. This is not surprising, because in the momentum range 
$0.3 < p_T < 1.5$~\GeVc chosen for ALICE analysis \cite{alice_2013}, 
the majority of particles is soft. It is worth noting that the 
balance function of the soft component has already been studied 
\cite{fu2011} in the framework of the \fastmc event generator 
\cite{Amelin:2006qe,Amelin:2007ic} and has been compared to the STAR 
measurements \cite{star_10}. The main results and conclusions of this 
study are the following. 
 
In the statistical model, directly produced particles, called 
``primordial", are generated independently, and there are no 
balancing charge correlations. As a result, the \fastmc balance 
function vanishes for the ``primordial'' particles. In this approach, 
the phase-space correlations between the final state particles arise 
only from the decays of hadronic resonances. Therefore, these 
correlations are determined merely by the kinematics of the decays. 
All charged particles falling into the considered pseudorapidity 
interval contribute to the denominator of Eq.~(\ref{bf}), whereas 
only those from the resonance decays, i.e., correlated pairs, 
contribute to the numerator when one calculates the balance function 
for the soft component. At RHIC energies, for instance, about a 
quarter of the observed pions at the freeze-out are primordial ones, 
while the remainder are produced via resonance decays.
 
The balance functions for the soft component in terms of relative 
pseudorapidity, $\Delta\eta$, depend on the maximum transverse flow 
velocity and the thermal freeze-out temperature. The option with 
large transverse flow and low thermal freeze-out temperature 
generally produces a relatively smaller width of the balance 
function than does the higher thermal freeze-out temperature option. 
The width of the balance function is inversely proportional to the 
strength of the transverse flow. The balance functions calculated 
in the framework of the \fastmc generator are usually lower than 
those measured in central Au+Au collisions at center-of-mass energy 
200 GeV per nucleon pair \cite{fu2011}, indicating that there are 
other sources of charge correlations besides resonance decays. 
However, the widths of \fastmc balance function are rather close to 
those measured in central Au+Au collisions, indicating that an 
important source of the correlation between the opposite-charge 
particles is the decay of resonances. 

\begin{figure}[htpb]
\includegraphics[width=0.49\textwidth]{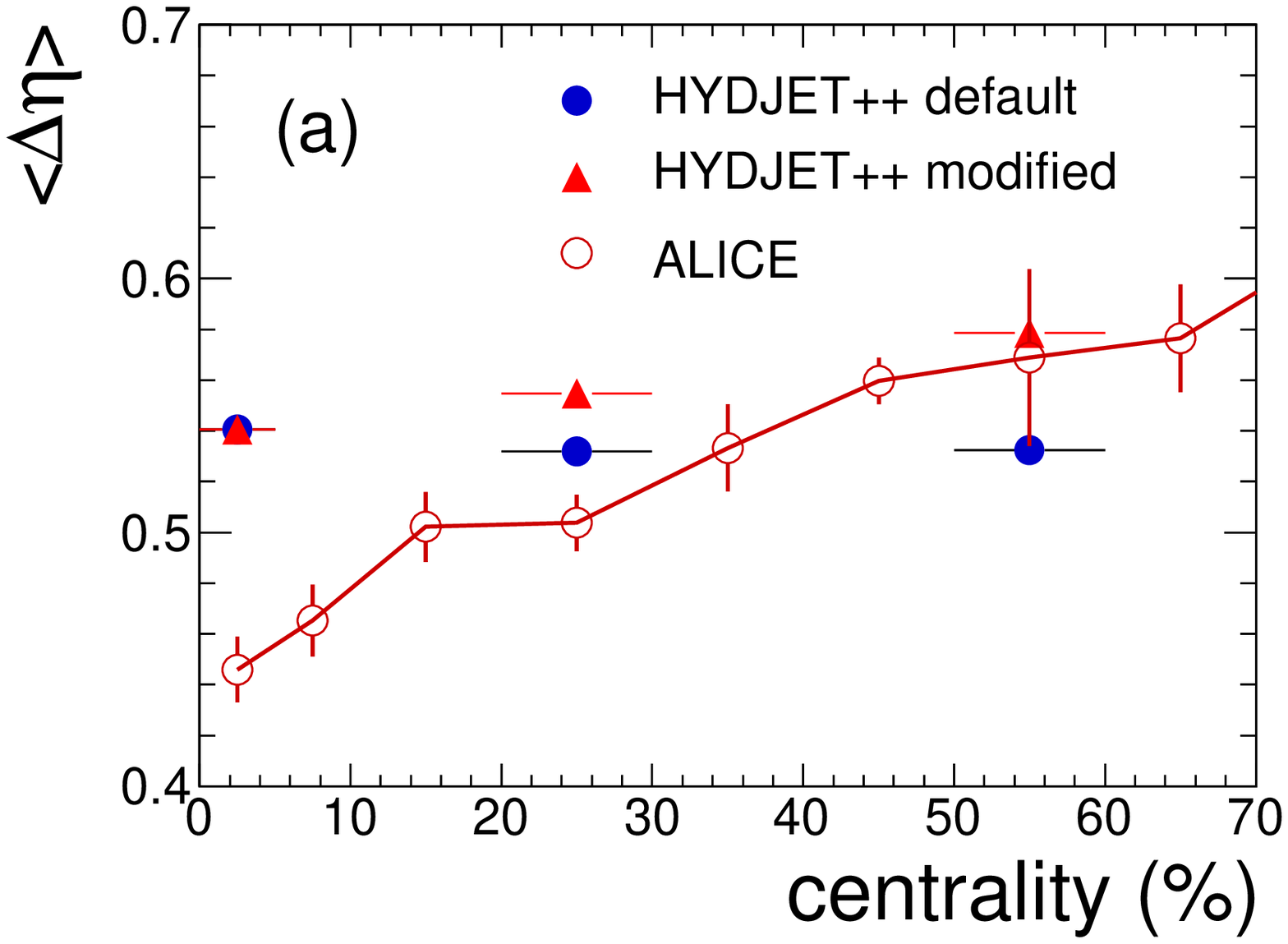}
\includegraphics[width=0.49\textwidth]{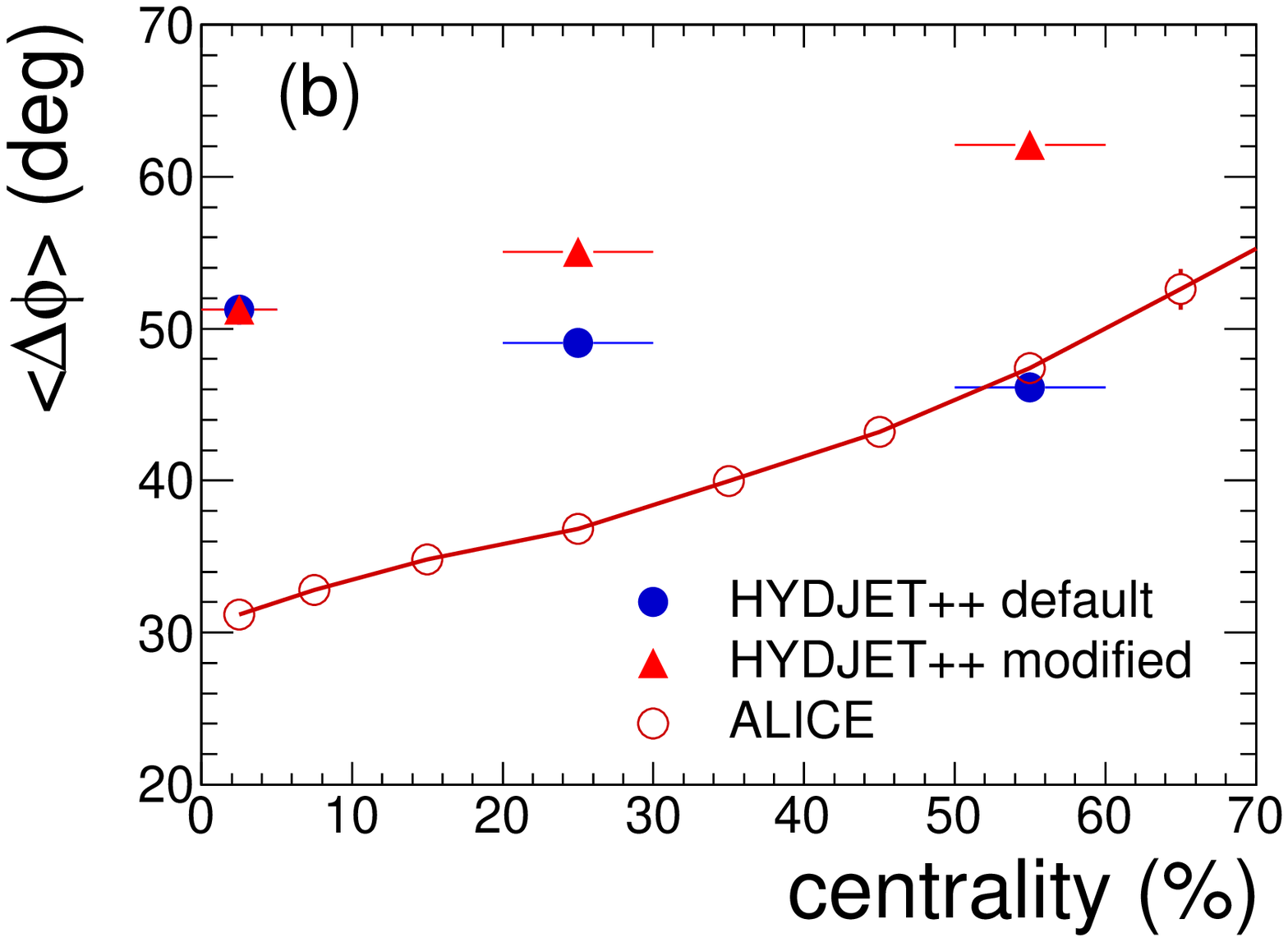}
\caption{(Color online) 
Centrality dependence of the width of the balance function of 
charged hadrons in Pb+Pb collisions at $\sqrtsnn = 2.76$~TeV for the 
correlations studied in terms of (a) relative pseudorapidity $\langle 
\Delta\eta\rangle$ and (b) relative angle $\langle\Delta\phi\rangle$, 
respectively. 
The model calculations with default (full circles) and modified 
(triangles) versions of \hydjet++ are compared to the ALICE data 
(open circles) from \protect\cite{alice_2013}. Lines are drawn to 
guide the eye.
}
\label{fig_alice2}
\end{figure}

The balance function in Fig.~\ref{fig_alice1} for the soft hadrons 
coming from the hard component (for $0.3< p_T <1.5$ \GeVc ) is higher 
than that for the soft component, and its width is broader than that 
of the soft component. This means that the charge correlations 
of the jet particles are weaker than those from the decay of 
resonances, which can be explained by the fact that the number of 
parton decays in the parton cascade during the jet fragmentation is 
larger than the number of consecutive resonance decays. Each 
subsequent decay makes the charge correlations weaker. The visible 
difference between the widths of soft and hard components opens some 
room for effectively reproducing the experimentally observed 
centrality dependence of the balance function widths. As a matter of 
fact, at the default parameters of \hydjet++ model the widths of the 
balance function reveal practically no centrality dependence, as 
shown in Fig. \ref{fig_alice2}. The widths are calculated in the 
entire interval where the balance function is measured, i.e., 
$|\Delta\eta| < 1.6$ and $-\pi/2 <\Delta\varphi < 3\pi/2$.
The applicability of the \hydjet++ generator to very peripheral 
collisions with centrality greater than 60\% is doubtful due to 
a number of physical assumptions and approximations made in the 
model. 

The centrality independence of the default model calculations 
indicates that some enhancement of the relative contribution from 
the jet part to the balance function is needed to reproduce the 
experimentally observed increase of the balance function widths in 
peripheral Pb+Pb collisions. Indeed, the mean ``soft'' and ``hard'' 
multiplicities depend on the centrality in different ways: They are 
roughly proportional to the mean number of participant nucleons 
$\langle N_{\rm part}(b) \rangle$ and the mean number of binary $NN$ 
subcollisions $\langle N_{\rm bin}(b) \rangle $ at a given impact 
parameter $b$, respectively. The relative contribution of the soft 
and hard parts to the total event multiplicity is fixed through the 
centrality dependence of the pseudorapidity hadron spectra 
$dN/d\eta$. The corresponding contributions from the hydro and jet 
parts are mainly determined by the two input parameters: 
$\mu_{\pi}^{\rm eff th}$, which is the (effective) chemical 
potential of positively charged pions at thermal freeze-out, and 
the minimum transverse momentum transfer of hard parton-parton 
scattering, $p_T^{\rm min}$. 

\begin{figure}[htpb]
\includegraphics[width=0.49\textwidth]{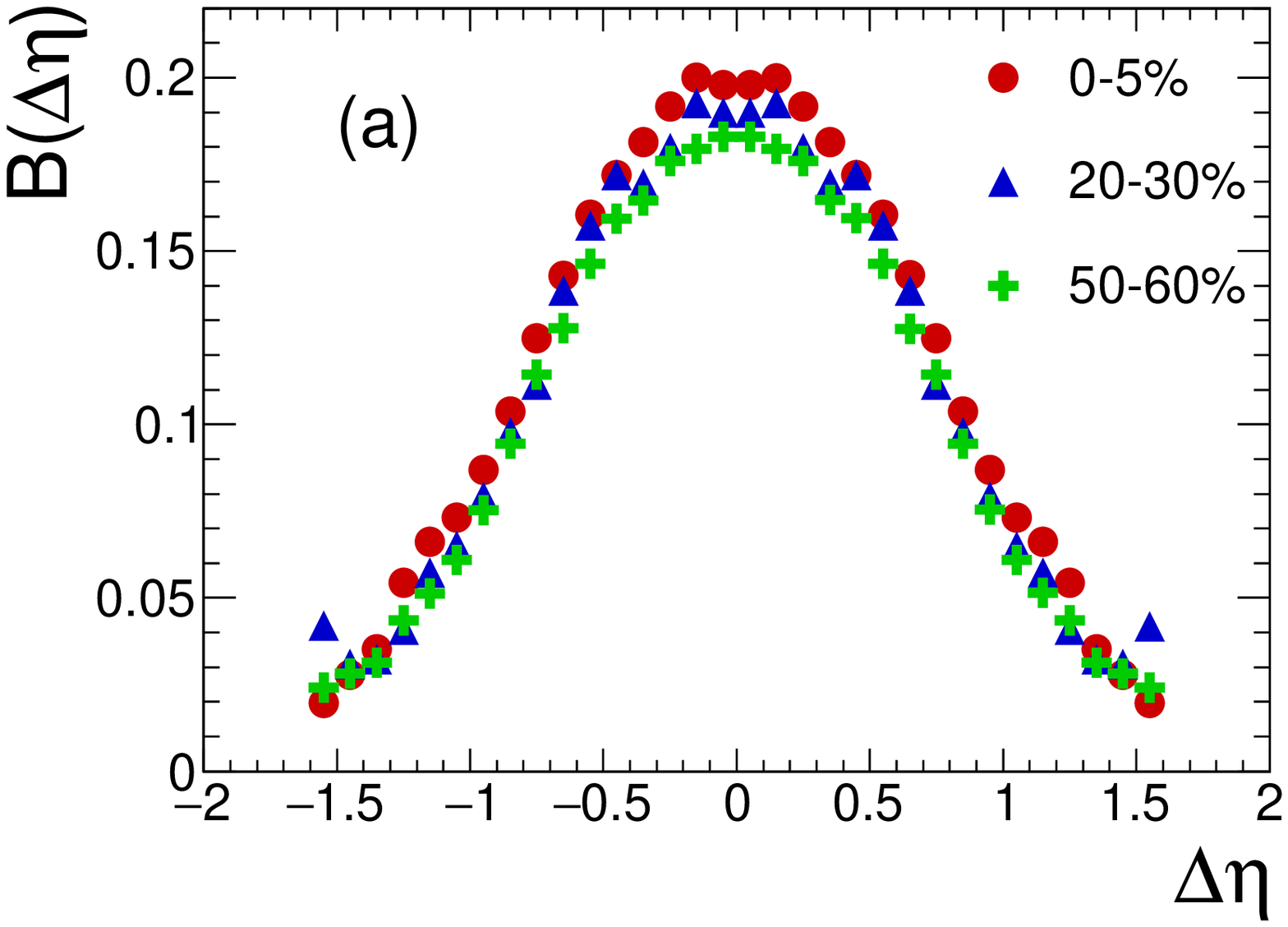}
\includegraphics[width=0.49\textwidth]{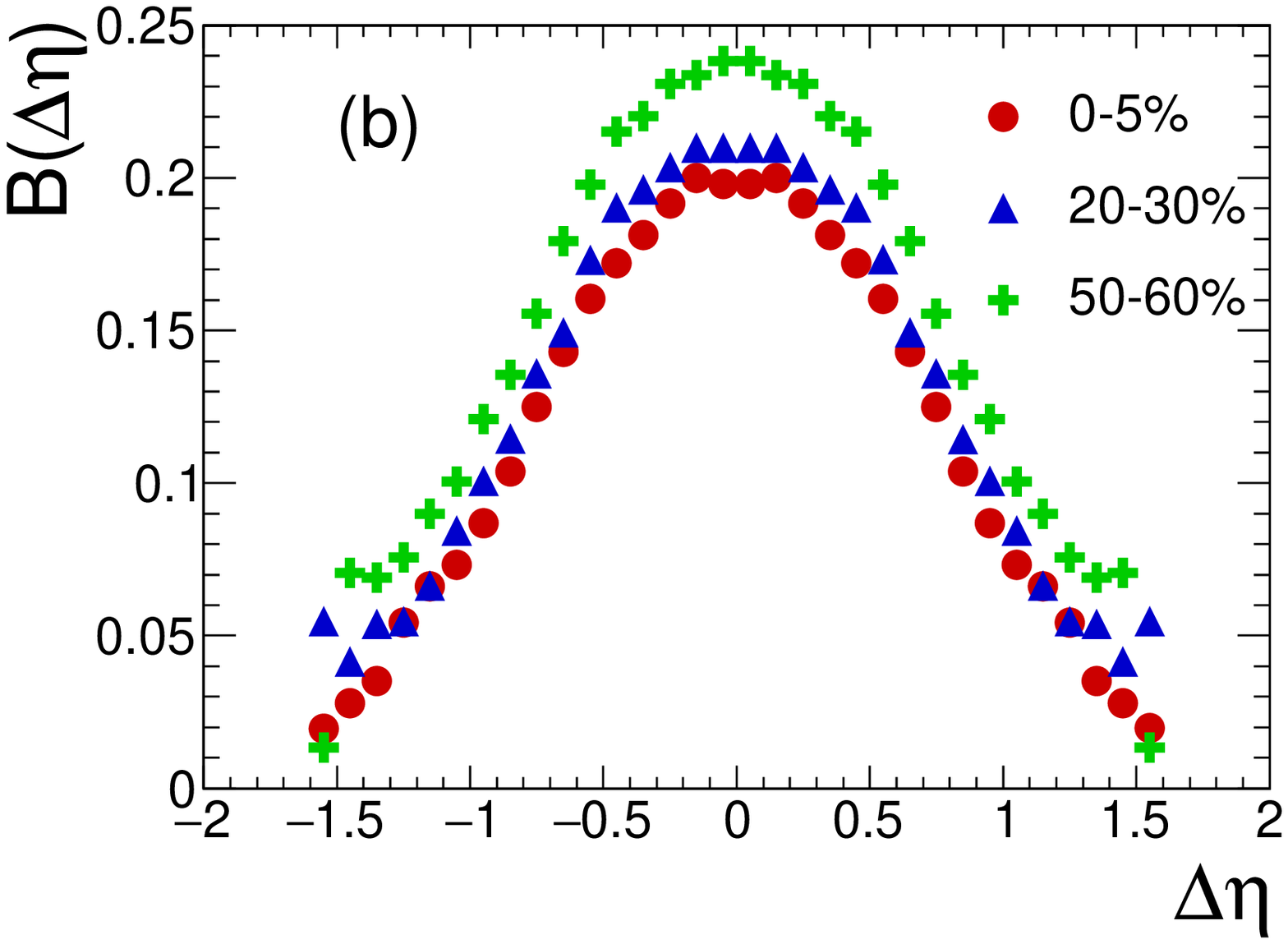}
\caption{(Color online)
Balance function $B(\Delta\eta)$ of charged hadrons in Pb+Pb 
collisions at $\sqrtsnn = 2.76$~TeV calculated with (a) \hydjet++ 
with the default parameters and (b) \hydjet++ with increased hard 
part and decreased soft part for three centralities, 0--5\% 
(circles), 20--30\% (triangles), and 50--60\% (crosses).
}
\label{fig_alice3}
\end{figure}

Besides, with the enhancement of the hard part for more peripheral 
collisions we decrease the contribution of the soft component. The 
contribution of the soft part to the total multiplicity varies with 
centrality because of its dependence on $\langle N_{\rm part}(b) 
\rangle$. We additionally decrease it by the factor $f_s$. The 
values of $p_T^{\rm min}$ and $f_s$, listed in Table~\ref{tab1}, 
are adjusted in such a way that the total spectra $dN/dp_T$ remains 
similar to that calculated with the default parameters 
$p_T^{\rm min} = 8.2$ \GeVc  and $f_s=1$ for a given centrality.
Note that the contribution of the hard component to the total 
particle multiplicity increases with event centrality in the 
\hydjet++ model. However, for the procedure described above such 
an increase is less pronounced than in the default HYDJET++ version. 
This is because the introduced additional enhancement of the hard 
part for more peripheral collisions is accompanied by a 
corresponding decrease in the contribution of the soft component in 
order to keep the total multiplicity for each event centrality the 
same as in the default version of \hydjet++.

\begin{table}
\caption{Parameters of the modified version of \hydjet++.
See text for details.}
\begin{center}
\begin{tabular}{ c | c | c }
\hline \hline
Centrality & $p_T^{\rm min}$, \GeVc  & $f_s$\\ 
\hline

 0-5\%   & 8.2  & 1      \\  

 20-30\% & 6.7  & 0.75   \\   

 50-60\% & 5.15 & 0.395  \\

\hline \hline

\end{tabular}
\end{center}
\label{tab1}
\end{table}

The widths for the modified version show the desired centrality 
dependence. However, our calculations are systematically higher 
than the experimental data. Moreover, the amplitudes of the modified 
balance functions shown in Fig.~\ref{fig_alice3} demonstrate a
centrality dependence opposite that of the default model 
calculations. These amplitudes increase with increasing impact 
parameter for calculations within the modified version, unlike
the default model results, in which the amplitudes decrease as in 
the experimental data \cite{alice_2013}. This discrepancy requires 
explanation. 

Since the relative contribution of the hard part increases in 
particular with the growth of the impact factor in the modified 
version, the width and amplitude of the total BF increase, because 
these values for the hard part are initially larger than those for 
the soft part. Somewhat unexpectedly, though only at first glance, 
there appears to be a simultaneous increase in both width and 
amplitude. This behavior is due to the fact that the BF normalization 
is not unique and is determined by hard, soft, and full multiplicity. 
In the case of Gaussian distributions with the same fixed 
normalization, the amplitudes always decrease with increasing width, 
in contrast to our situation with its own normalization for each 
centrality class. In the \hydjet++ model we have two 
independent sources of charge correlations, namely, the resonances 
with the width $\langle \Delta\eta\rangle _{\rm res}$ and the jets 
with $\langle \Delta\eta\rangle_{\rm jet}$. Note that 
$\langle \Delta\eta\rangle _{\rm res}$ is less than 
$\langle \Delta\eta\rangle_{\rm jet}$.
The resulting width has some intermediate value between 
$\langle \Delta\eta\rangle _{\rm res}$ and 
$\langle \Delta\eta\rangle_{\rm jet}$ depending on the relative 
contributions from these sources.
The widths $\langle \Delta\eta\rangle _{\rm res}$ and 
$\langle \Delta\eta\rangle_{\rm jet}$ are initially larger than 
$\langle \Delta\eta\rangle_{\rm exp}$; therefore, any resulting 
width from these sources will be larger than 
$\langle \Delta\eta\rangle_{\rm exp}$ as well. Our observation 
indicates that some additional source of charge correlation with a 
width less than $\langle \Delta\eta\rangle _{\rm res}$ is needed to 
reproduce the experimentally observed values in the interval of 
relatively low transverse momenta.

For higher transverse momentum regions it was found \cite{alice_epjc} 
that the balance function is increasingly narrow, thus indicating 
that the correlations become stronger. Moreover, the centrality 
dependence of the widths practically vanishes. We compare the widths 
calculated at the high transverse momenta in the intervals

(1) $2.0 < p_{T,\rm assoc} < 3.0 < p_{T,\rm trig} < 4.0$ \GeVc  and

\begin{figure}[htpb]
\includegraphics[width=0.49\textwidth]{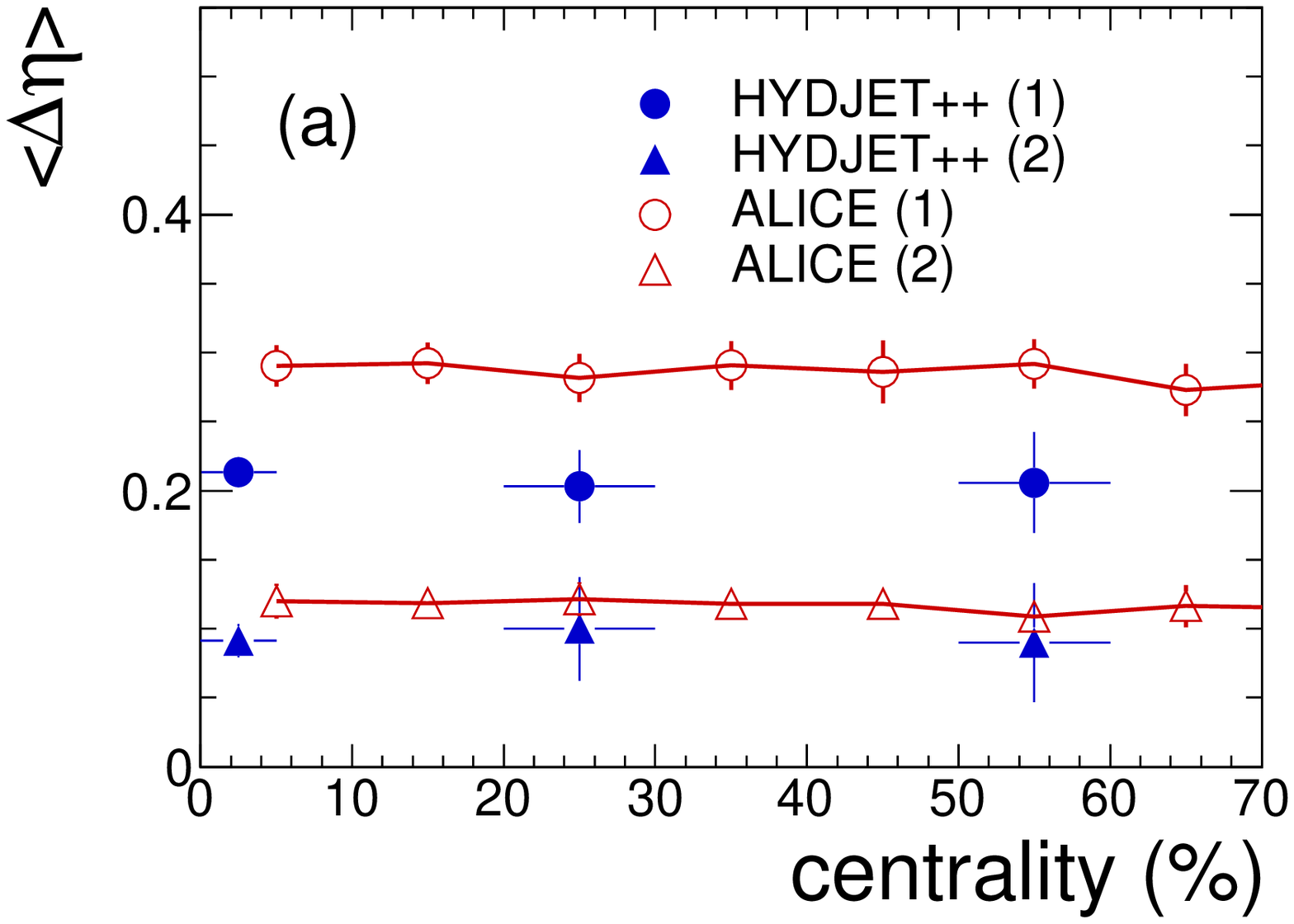}
\includegraphics[width=0.49\textwidth]{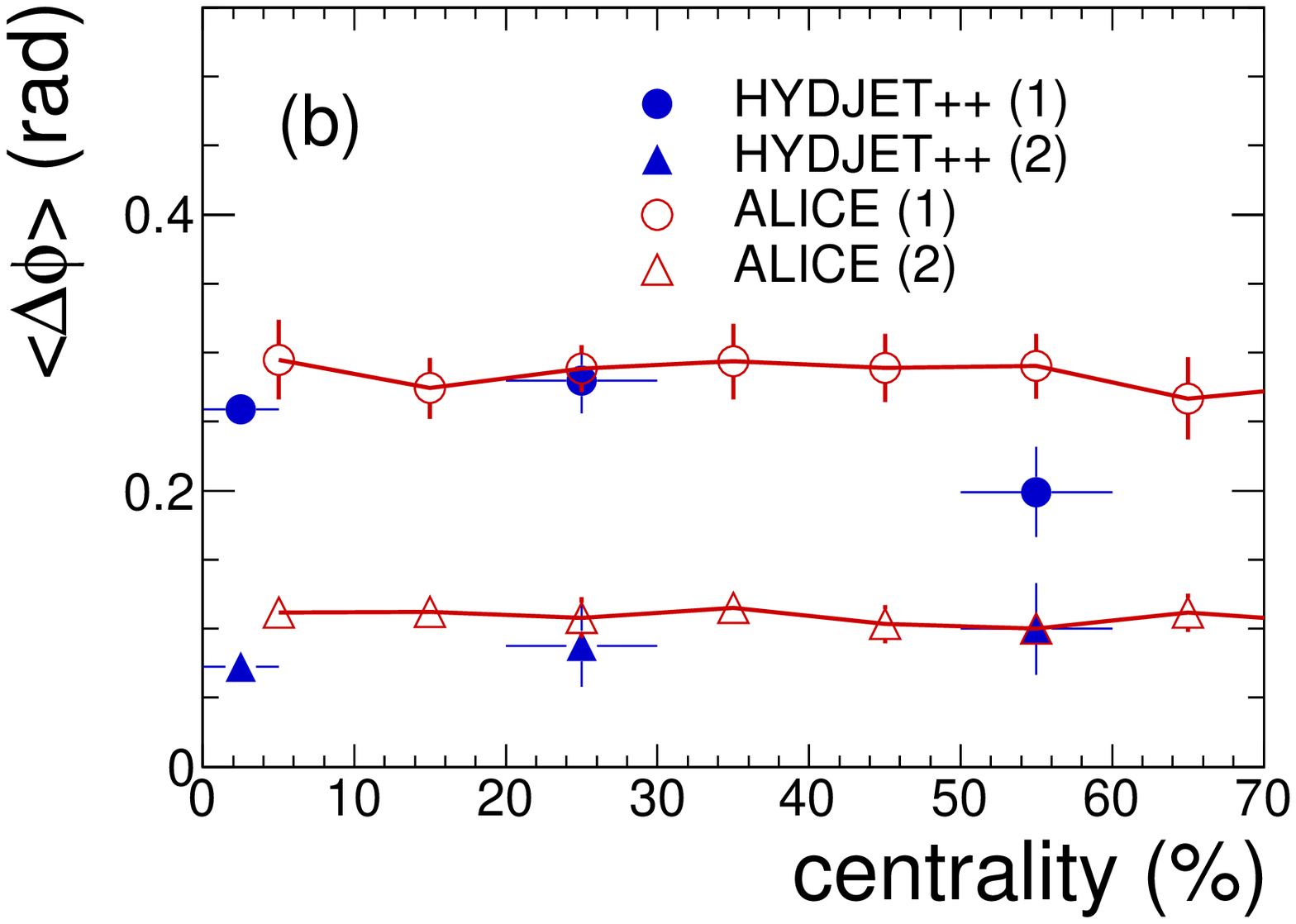}
	\caption{(Color online)
The same as Fig.~\ref{fig_alice2} but for two transverse momentum
intervals labeled in the text as (1) and (2). For interval (1)
\hydjet++ calculations and the data from \protect\cite{alice_2013} 
are denoted as full and open circles, respectively, whereas for 
interval (2) the calculations and the data are indicated as full 
and open triangles. Lines are drawn to guide the eye.
}
\label{fig_alice4}
\end{figure}

(2) $3.0 < p_{T,\rm assoc} < 8.0 < p_{T,\rm trig} < 15.0$ \GeVc

\noindent
with the data \cite{alice_epjc}. Here the widths were calculated 
in accordance with experimental calculations, not in the entire 
intervals of $\Delta\eta$, $\Delta\varphi$, but in the narrow 
interval. The balance function distributions are fitted to a sum 
of a Gaussian and a constant. The width is then calculated within 
$3 \sigma_{\rm Gauss}$. The comparison is displayed in 
Fig.~\ref{fig_alice4}. One can see in this figure that with 
increasing transverse momentum the default model results describe 
the experimental data much better, since in these transverse 
momentum intervals the contribution from the soft component is 
weaker. In practical terms we observe a transition to a single 
source of charge correlations, namely, charge correlations in jets 
for which exact charge conservation holds at each stage. The 
decrease of the width with increasing transverse momentum can also 
be explained as due to particles with higher transverse momenta 
being created closer to the beginning of parton cascade, where the 
charge correlations are stronger.

\section{Implementation of exact charge conservation in the model}
\label{sec_4}

Our investigation reveals that the interval of relatively low 
transverse momenta remains problematic in description of the 
charge balance function despite the attempts that have been made to
address it. It is a long-standing conceptual problem that the majority 
of soft particles are generated independently in the statistical model 
approach, and the charge correlations emerge merely as a result of the 
resonance decays. The widths $\langle \Delta\eta\rangle _{\rm res}$  
are larger than $\langle \Delta\eta\rangle_{\rm exp}$, and in 
accordance with the consideration above another source of charge 
correlation with width less than 
$\langle \Delta\eta\rangle _{\rm res}$ should be introduced to 
reproduce the experimentally observed values in the interval 
of relatively low $p_T$.

In order to take exact charge conservation into account, we 
modified the generation procedure of soft direct hadrons. In the 
default model version a number $N$ of particles with specified 
coordinates and momenta are generated in each event. In the 
modified version the following procedure is realized. For the sake
of simplicity, we consider here the electrically neutral system
similar to the midrapidity part of the fireball produced in heavy-ion 
collisions at LHC energies. First, half of all charged particles are 
removed randomly in every event, whereas all neutral particles remain
unaffected. Then, for each particle with a given charge, an 
antiparticle with opposite charge is added to the particle spectrum. 
For instance, $\pi^-$ should be added to $\pi^+$, and vice versa.
The coordinates and transverse momentum of each new particle are 
assigned the same values as the corresponding original particle, 
ensuring local charge conservation. The pseudorapidities and the 
azimuthal angles of these new particles are distributed around the 
pseudorapidities and the azimuthal angles of the corresponding 
original particles. The Gaussian distributions are used

\begin{equation}
\displaystyle 
P(x) = \frac{1}{\sqrt{2 \pi} \sigma_x} \times 
       \exp{ \left[ - \frac{(x - x_0)^2}{2 \sigma_x^2} \right] }
\label{Gauss}
\end{equation}

Here $x = \{\eta ; \varphi \}$, and the dispersions $\sigma_x$ of 
the distributions are new parameters of the model. They should yield
widths of the charge balance function less than 
$\langle \Delta\eta\rangle _{\rm res}$ and 
$\langle \Delta\varphi\rangle _{\rm res}$ to reproduce the 
experimental data in the most central collisions. As a result, 
equal numbers of particles with positive and negative charges will 
be obtained in each newly generated event. (As a matter of fact, the 
pairs of particles and antiparticles with correlated pseudorapidity 
and azimuthal angle are generated in a similar way in other models.)
Finally, the spectra of the newly generated direct hadrons remain 
almost unchanged, but their balance function will be nonzero with a 
some finite width, in contrast to the original procedure with the 
formally infinite width of soft direct hadrons, because we have no 
charge correlations at all due to the independent generation of 
particles. The dispersion of Gaussian distributions should be fitted 
to reproduce the centrality dependence of widths in the intervals of 
relatively low transverse momenta.

\begin{figure}[htpb]
\includegraphics[width=0.49\textwidth]{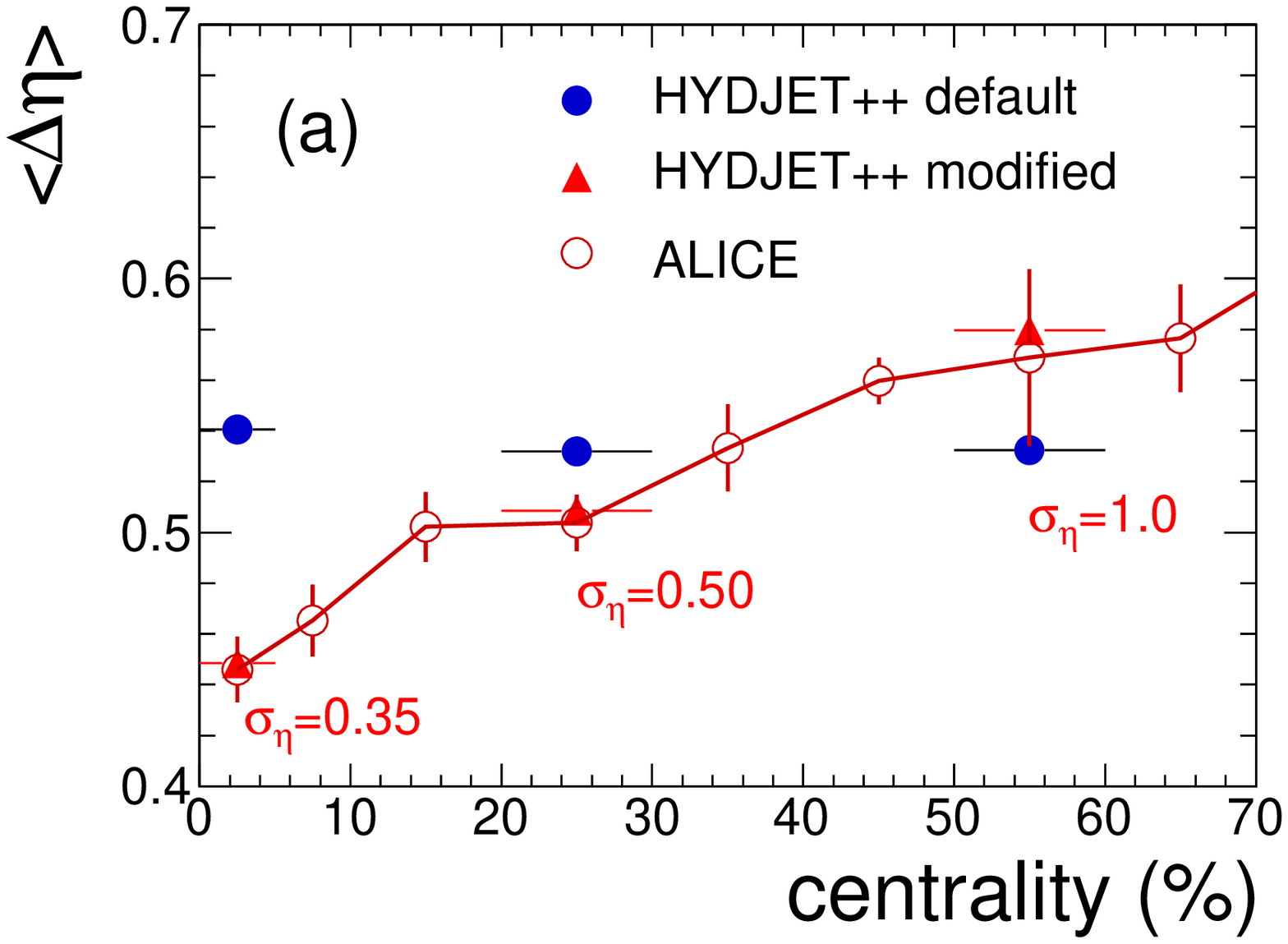}
\includegraphics[width=0.49\textwidth]{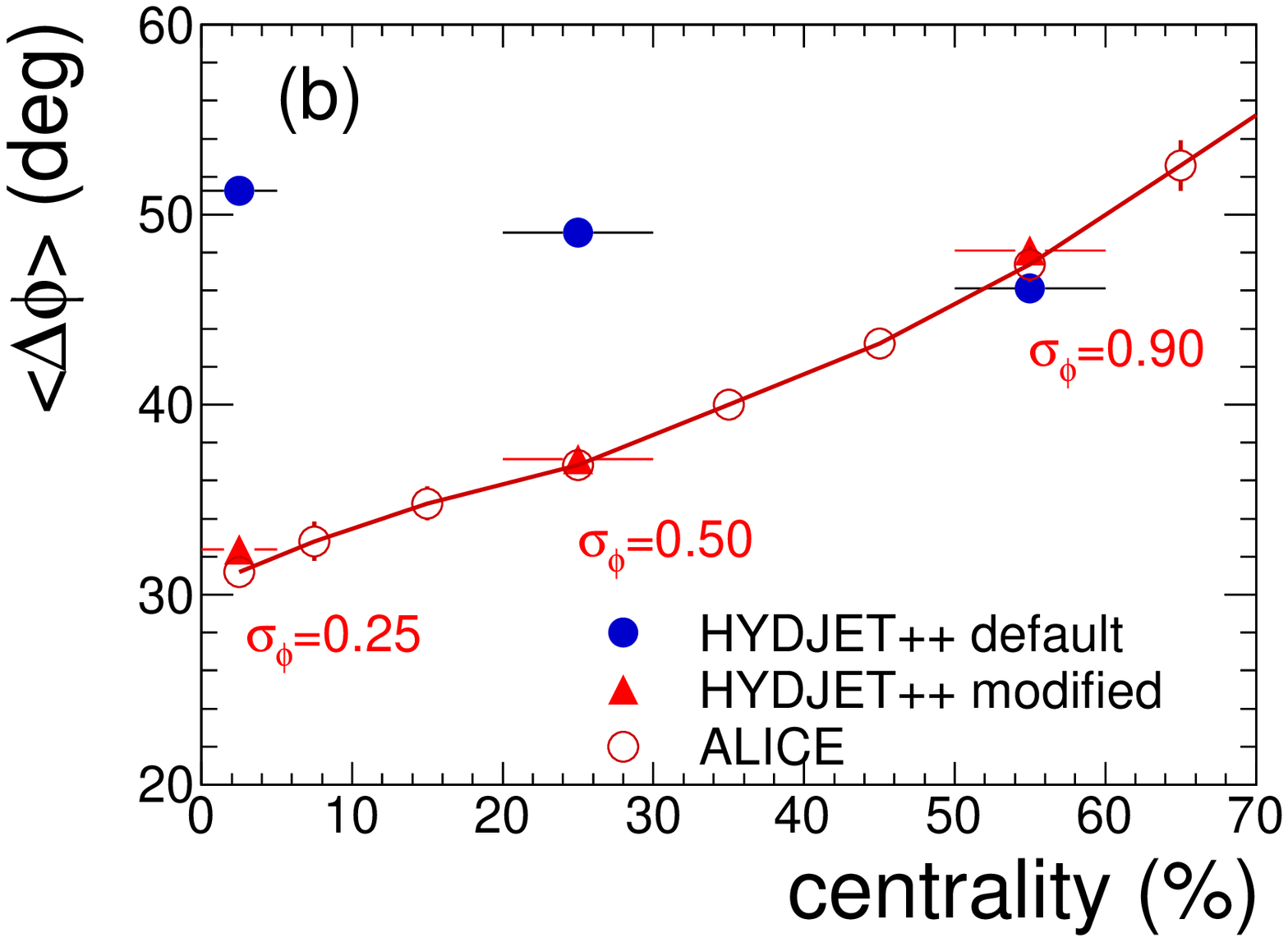}
	\caption{(Color online)
The same as Fig.~\protect\ref{fig_alice2} but for calculations with
a modified version of \hydjet++ (triangles) incorporating exact 
charge conservation. Full circles denote the default version 
calculations and open circles indicate the experimental data. 
Lines are drawn to guide the eye.
}
\label{fig_alice5}
\end{figure}

The results of our fitting procedure are shown in Fig.~\ref{fig_alice5}. 
The model calculations reproduce the experimentally observed centrality
dependence of widths rather well if the dispersions of Gaussian 
distributions increase with the growth of the impact parameter. This 
indicates that the charge correlations of direct hadrons become weaker 
in more peripheral collisions than central ones. 

To reveal the physical mechanism that provides a shorter correlation 
length than that given by resonance decays and jet fragmentation at the 
freeze-out stage, a full description of system evolution is necessary. 
In the statistical and hydrodynamical models, particles are produced at 
the freeze-out hypersurfaces. In these models part of the information 
about early system dynamics is encoded in the parameters of the 
emission source of soft hadrons; for example, the strength and 
direction of the elliptic flow are governed by two parameters 
characterizing the momentum and spatial anisotropy of the emission 
source. Similarly, the charge correlations of the hadrons directly 
produced at the freeze-out stage are encoded using the dispersions of 
Gaussian distributions. An explanation why these dispersions should 
increase with an increase of the impact parameter is as follows. The 
number of characteristic elementary volumes, i.e., the independent 
particle sources, within which the charge is explicitly conserved, 
decreases with increasing impact parameter, since the area of the 
nuclear overlap region becomes smaller. This means that the 
fluctuations become stronger \cite{Eyyubova_2021}, thus destroying 
the charge correlation in general.  
The fluctuation centrality dependence \cite{Eyyubova_2021} of the 
dispersions can be presented in a form
\begin{equation}
\displaystyle
 \sigma_x(C) = \sigma_x (C_0) 
	\sqrt{\frac{1 - C_0^{1/2}}{1 - C^{1/2}}} \ ,
\label{fluct_centr}
\end{equation}
where $\sigma_x (C_0)$ should be fixed at some centrality $C_0$, e.g., 
for the most central collisions $C_0=0$ and $\sigma_{\eta} (0)= 0.35$,
as follows from our fitting procedure. After that, other dispersions 
can be calculated. Their values are very close to the fitting values 
presented in Fig.~\ref{fig_alice5}. Thus, the simple fluctuation formula 
given by Eq.(\ref{fluct_centr}) can be used as a good approximation 
instead of the fitting procedure in every centrality bin.

\begin{figure}[htpb]
\includegraphics[width=0.50\textwidth]{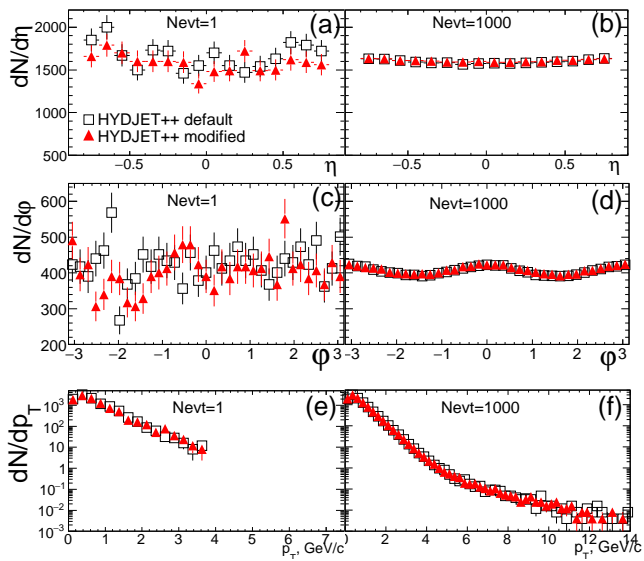}
	\caption{(Color online)
Charged particle multiplicity versus pseudorapidity (a),(b),
azimuthal angle (c),(d), and transverse momentum (e),(f) calculated
in the default (open squares) and modified (full triangles)
versions of \hydjet++ for one event (left column) and one
thousand events (right column). 
}
\label{spectra}
\end{figure}

The total energy-momentum of the new system and the original are
slightly different in each single event, but in the statistical a
pproach, the total energy-momentum, together with the number of 
particles, varies (fluctuates) from event to event; only their averages 
and moments make sense. The large number of particles in each event 
during the random selection procedure ensures that the average of the 
considered values remain unchanged in case of large statistics of 
generated events. To demonstrate this explicitly, we provide in 
Fig.~\ref{spectra} the multiplicity distributions of charged particles 
versus pseudorapidity [Fig.~\ref{spectra}(a),(b)], azimuthal angle 
[Fig.~\ref{spectra}(c),(d)], and transverse momentum 
[Fig.~\ref{spectra}(e),(f)], calculated in the default and modified 
versions of \hydjet++ for one event (left column) and for
one thousand events (right column). It is clear that despite some
differences between the spectra for a single event, the $dN/d \eta$, 
$dN/d \phi$, and $dN/d p_T$ distributions calculated for 1000 events 
with the modified and default versions coincide.

Here it is worth mentioning a recently developed method 
\cite{Oliinychenko} to ensure conservation laws for each sampled 
configuration in spatially compact regions, or patches, at the 
freeze-out stage. This method allows one also to study the correlation 
effects sensitive to the patch size as a fitting parameter. Our 
procedure is considerably simpler for the realization in Monte Carlo 
event generators and does not imply any modification of single 
particle spectra or any additional tuning of other free parameters 
of the model.

\section{Conclusions}
\label{sec_5}

The phenomenological analysis of the charge balance function in Pb+Pb 
collisions at center-of-mass energy  2.76 TeV per nucleon pair has been 
performed within the two-component \hydjet++ model. It is shown that 
the experimentally observed increase of the balance function width with 
increasing impact parameter in the relatively low transverse momentum 
interval can be qualitatively reproduced  by the relative enhancement of
the contribution from the hard component in peripheral Pb+Pb collisions.
However, the centrality dependence of the CBF magnitude shows an 
opposite tendency to the experimental one. 
A fully adequate description of the balance function in this interval
assumes modification of the essential model through the explicit 
inclusion of charge conservation in a statistical approach. This 
procedure has been implemented for the first time in Monte Carlo event 
generators of such a kind.
With increasing transverse momentum the default model results describe 
the experimental data much better, because in these transverse momentum 
intervals the contribution from the soft component of the model is 
weakened. In practical terms, we find a transition to a single source 
of charge correlations, namely, the charge correlations in jets for 
which exact charge conservation holds at each stage.

\section*{Acknowledgments}
{\it Fruitful discussions with A.V.~Belyaev, L.V.~Bravina, and 
A.I.~Demyanov are gratefully acknowledged.}

\end{document}